\def\be{\begin{equation}}
\def\ee{\end{equation}}
\def\ba{\begin{array}}
\def\ea{\end{array}}

\documentclass[aps,amsmath,amssymb,amsfonts,showpacs]{revtex4}
\usepackage{graphicx}
\usepackage{epstopdf}
\def\qed{\leavevmode\unskip\penalty9999 \hbox{}\nobreak\hfill
     \quad\hbox{\leavevmode  \hbox to.77778em{%
               \hfil\vrule   \vbox to.675em%
               {\hrule width.6em\vfil\hrule}\vrule\hfil}}
     \par\vskip3pt}

\newtheorem{theorem}{Theorem}
\newtheorem{lemma}{Lemma}

\begin{document}
\title{ Quantifying quantum coherence based on the Tsallis relative operator entropy}
\author{Meng-Li Guo$^{1}$}
\author{Zhi-Xiang Jin$^{2}$}
\author{Bo Li$^{3}$}
\email{libobeijing2008@163.com.}
\author{Bin Hu$^{1}$}
\email{bhu@ecit.cn.}
\author{Shao-Ming Fei$^{4,5}$}

\affiliation{$^1$Department of Mathematics, East China University of Technology, Nanchang 330013, China\\
$^2$School of Physics, University of Chinese Academy of Sciences, Yuquan Road 19A, Beijing 100049, China\\
$^3$School of Mathematics and Computer science, Shangrao Normal University, Shangrao 334001, China\\
$^4$School of Mathematical Sciences, Capital Normal University, Beijing 100048, China\\
$^5$Max-Planck-Institute for Mathematics in the Sciences, 04103, Leipzig, Germany}

\begin{abstract}
Coherence is a fundamental ingredient in quantum physics and a key resource in quantum information processing.
The quantification of quantum coherence is of great importance.
We present a family of coherence quantifiers based on the Tsallis relative operator entropy.
Shannon inequality and its reverse one in Hilbert space operators derived by Furuta [Linear Algebra
Appl. 381 (2004) 219] are extended in terms of the parameter of the Tsallis relative operator
entropy
These quantifiers are shown to satisfy
all the standard criteria for a well-defined measure of coherence, and include some existing coherence measures as special cases.
Detailed examples are given to show the relations among the measures of quantum coherence.
\end{abstract}

\maketitle

\section{Introduction}
Quantum coherence is one of the most fundamental physical resources in quantum mechanics,
which can be used in quantum optics \cite{c1}, quantum information and quantum computation \cite{c2},
thermodynamics \cite{c3,c4} and low temperature thermodynamics \cite{c5,c6,c7,c8}.
Coherent quantification is one of the most important ingredient not only in quantum theory but also in practical applications.
Recently, resource theory of coherence based on positive operator valued measurement (POVM) has been studied in \cite{nc8,nc9,nc10}. This approach provide us to understand coherence in more fundamental way as POVMs are the most general kind of quantum measurements.
In Ref. \cite{c9}, the author established a consistent framework of resource theory to quantify coherence.
In this theory, the coherence describes the superposition of quantum states relative to fixed orthogonal bases.
Since then, a lot of work has been done to enrich this theory \cite{c10,cq1,c11,X2,JX,c12}.
This framework has some important limitations on the measurement of coherence.
Different coherence measures may reflect different physical aspects of quantum systems \cite{c13,J1,Jin,PR37,PRA6,c14,c15}.

Let $\mathcal{H}$ be a finite dimensional Hilbert space with an orthogonal basis $\{|i\rangle\}^d_{i=1}$.
In this basis, the diagonal density matrices are free states \cite{c16},
which are also called incoherent states. We label the set of incoherent quantum states as $\mathcal{I}$,
\begin{equation}\nonumber
\mathcal{I}=\{\sigma ~|~ \sigma=\sum^d_{i=1}\lambda_i |i\rangle\langle i|\}.
\end{equation}

Free operation in coherence theory is a completely positive and trace preservation (CPTP) mapping,
which admits an incoherent Kraus representation.
Namely, there always exists a set of Kraus operators $\{K_i\}$ such that
\begin{equation}\nonumber
\frac{K_i\sigma K^{\dagger}_i}{\mathrm{Tr}K_i\sigma K^{\dagger}_i}\in\mathcal{I},
\end{equation}
for each $i$ and any incoherent state $\sigma$. These operations are also called incoherent operations and we label them by $\Phi$.

Similar to the quantification of entanglement \cite{c17,c18,c19,c20},
any measure of coherence $C$ should satisfy the following axioms \cite{c9}:

$(C_1)$ Faithfulness: $C(\rho)\geq0,$ for all quantum states $\rho$, and $C(\rho)=0$ if and only if $\rho\in \mathcal{I}$;

$(C_2)$ Monotonicity: $C$ does not increase under incoherent completely positive and trace preserving maps (ICPTP) $\Phi$, i.e.,
\begin{equation}\nonumber
C(\Phi(\rho))\leq C(\rho);
\end{equation}

$(C_3)$ Strong monotonicity: $C$ does not increase on average under selective incoherent operations, i.e.,
\begin{equation}\nonumber
\sum_ip_iC(\sigma_i)\le C(\rho),
\end{equation}
with probabilities $p_i=\mathrm{Tr}K_i\rho K^{\dagger}_i$,
post-measurement states $\sigma_i= K_i\rho K^{\dagger}_i/p_i$,
and incoherent operators $K_i$;

$(C_4)$ Convexity: Non-increasing under mixing of quantum states, i.e.,
\begin{equation}\nonumber
\sum_ip_iC(\rho_i)\ge C(\sum_ip_i\rho_i),
\end{equation}
for any set of states $\{\rho_i\}$ and $p_i\ge0$ with $\sum_i p_i=1$.

In \cite{c94} the authors show that the conditions $(C_3)$ and $(C_4)$ are equivalent to the following additivity of coherence for block-diagonal states,

$(C_5)$ \begin{equation*}
C(p\rho_1\oplus(1-p)\rho_2)=pC(\rho_1)+(1-p)C(\rho_2),
\end{equation*}
for any $p\in[0,1]$, $\rho_i\in\varepsilon(\mathcal{H}_i)$, $i=1, 2$,
and $p\rho_1\oplus(1-p)\rho_2\in\varepsilon(\mathcal{H}_1\oplus \mathcal{H}_2),$
where $\varepsilon(\mathcal{H})$ denotes the set of density matrices on the Hilbert space $\mathcal{H}$.

Other frameworks for quantifying coherence have been further investigated \cite{c21,c22,c23}.
So far, various quantities have been proposed to serve as a coherence quantifier, however the available candidates are still quite limited. Up to now, many coherence measures have been proposed based on different applications and backgrounds,
such as the relative entropy of coherence \cite{c9}, the $l_1$ norm of coherence \cite{c9}, geometric coherence \cite{c24}, coherence
measures based on Tsallis relative entropy \cite{c25,nc25,c26} and so on.
The Tsallis relative entropy lays the foundation to the non-extensive thermo-statistics and have important applications in the information theory.
But it was shown to violate the strong monotonicity, even though it can unambiguously distinguish the coherent and the incoherent
states with the monotonicity.
Here we establish a class of coherence quantifiers which are closely  related to the Tsallis relative $\alpha$ entropy. It proves that this family of quantifiers satisfy all the standard criteria and particularly cover several typical coherence measures.
 Therefore, it is  important  to study the properties of the coherence measures based on the Tsallis relative entropy, as well as studying the properties before taking a trace, that is, the Tsallis relative operator entropy, which is a parametric extension of the relative operator entropy.

In this paper, we provide a class of coherence measures based on the Tsallis relative operator entropy.
Tsallis relative operator entropy was defined as a parametric extension of relative operator entropy \cite{c27}.
It is meaningful to study the properties of Tsallis relative operator entropy for the development of the noncommutative statistical physics and nonadditive quantum information theory, therefore,
we think it is indispensable to study the coherence of Tsallis relative operator entropy as a versatile resource for quantum information protocols. This paper is organized as follows. We first introduce the coherence measure and  the Tsallis relative operator entropy.
Then we present the family of coherence quantifier satisfy all the standard criteria for  well-defined measures of coherence,
then we study the maximal coherence and  include some existing coherence measures as special cases.
Detailed examples are given to show the relations among the measures of quantum coherence.
Finally, we finish the paper by the conclusion.

\section{Coherence quantification}
We first recall the Tsallis relative operator entropy.
As we all know, a bounded linear operator $T$ on a Hilbert space $H$ is said to be positive
(denoted by $T\geq0$) if the inner product $(T x, x)\geq0$ for all $ x\in H$, and an operator $T$ is said to be strictly
positive (denoted by $T>0$ ) if $T$ is invertible and positive.
As ``T'' is a linear and Hermitian operator we have that range $T$ is simply a subspace spanned by all eigenvectors of $T$ belonging
to nonzero eigenvalues.
The Tsallis relative operator entropy is defined by
\begin{equation}\label{n2}
T_{q}(\rho||\sigma)
=\frac{\rho^{\frac{1}{2}}\left(\rho^{-\frac{1}{2}}\sigma \rho^{-\frac{1}{2}}\right)^{1-q}\rho^{\frac{1}{2}}-\rho}{1-q},
\end{equation}
for arbitrary two invertible positive operators $\rho$ and $\sigma$ on Hilbert space, and any real number $q\in [0,1)$.
For convenience, one writes $T_{q}(\rho||\sigma)$ as \cite{clnx},
\begin{equation*}
T_{q}(\rho||\sigma)
=\rho^{\frac{1}{2}}\ln_{1-q}(\rho^{-\frac{1}{2}}\sigma \rho^{-\frac{1}{2}}) \rho^{\frac{1}{2}},
\end{equation*}
where $\ln_{1-q}X\equiv\frac{X^{1-q}-1}{1-q}$ for the positive operator $X=\rho^{-\frac{1}{2}}\sigma \rho^{-\frac{1}{2}}$.
As for any strictly positive real number $x$ and $q\in [0,1)$, the following inequalities hold:
\begin{equation}\label{bby}
1-\frac{1}{x}\leq \ln_{1-q} x\leq x-1,
\end{equation}
we have the following conclusion.

\begin{lemma}
For any positive invertible operators $\rho$ and $\sigma$ and $q\in [0,1)$,
\begin{equation*}
\rho-\rho\sigma^{-1}\rho\leq T_{q}(\rho||\sigma)\leq \sigma-\rho.
\end{equation*}
Moreover, $T_{q}(\rho||\sigma)=0$ if and only if $\rho=\sigma$.
\end{lemma}

{\sf [Proof]} According to (\ref {bby}), we have
\begin{equation*}
\mathrm{I}-\rho^{\frac{1}{2}}\sigma^{-1} \rho^{\frac{1}{2}}\leq \ln_{1-q}(\rho^{-\frac{1}{2}}\sigma \rho^{-\frac{1}{2}})\leq \mathrm{-I}+\rho^{-\frac{1}{2}}\sigma \rho^{-\frac{1}{2}}.
\end{equation*}
Multiplying $\rho^{\frac{1}{2}}$ on both sides of the terms in above inequality, one gets
\begin{equation*}
\rho^{\frac{1}{2}}(\mathrm{I}-\rho^{\frac{1}{2}}\sigma^{-1} \rho^{\frac{1}{2}})\rho^{\frac{1}{2}}\leq \rho^{\frac{1}{2}} \ln_{1-q}(\rho^{-\frac{1}{2}}\sigma \rho^{-\frac{1}{2}}) \rho^{\frac{1}{2}}\leq \rho^{\frac{1}{2}}(\mathrm{-I}+\rho^{-\frac{1}{2}}\sigma \rho^{-\frac{1}{2}})\rho^{\frac{1}{2}}.
\end{equation*}
Hence, $\rho-\rho\sigma^{-1}\rho\leq T_{q}(\rho||\sigma)\leq \sigma-\rho$.
Moreover, suppose $T_{q}(\rho||\sigma)=0$, then
$\rho-\rho\sigma^{-1}\rho\leq 0 \leq \sigma-\rho$,
which implies that $\rho\geq \sigma$ and $\rho\leq \sigma$, namely, $\rho=\sigma$. If $\rho=\sigma$, one can easily verify that $T_{q}(\rho||\sigma)=0$.
$\Box$

In addition, $T_q(\rho||\sigma)$ satisfies the following properties \cite{clnx}:

(I) (homogeneity) $T_q (\alpha \rho||\alpha \sigma) = \alpha T_q(\rho||\sigma)$ for any positive number $\alpha$.

(II) (monotonicity) If $\sigma \leq \tau$, then $T_q(\rho||\sigma)\leq T_q(\rho||\tau)$.

(III) (superadditivity) $T_q(\rho_1+\rho_2||\sigma_1+\sigma_2)\geq T_q(\rho_1||\sigma_1)+T_q(\rho_2||\sigma_2)$.

(IV) (joint concavity) $T_q(\alpha \rho_1+\beta \rho_2||\alpha\sigma_1+\beta \sigma_2)\geq \alpha T_q(\rho_1||\sigma_1)+\beta T_q(\rho_2||\sigma_2)$.

(V) For any unitary operator $T_q(U \rho U^{\dagger }||U \sigma U^{\dagger }) =T_q(\rho||\sigma) $.

(VI) For a unital positive linear map $\Phi$ from the set of the bounded linear operators on Hilbert space to itself, one has
$\Phi(T_q(\rho||\sigma))\leq T_q(\Phi(\rho)||\Phi(\sigma))$.

The Tsallis relative $\alpha$ entropy is a special case of the quantum $f$-divergences \cite{c26}.
Moreover, in order to make our definition correspond to the definition of the relative operator entropy defined, we change the sign of the original Tsallis relative $\alpha$ entropy.
For two density matrices $\rho $ and $\sigma $, the Tsallis relative $\alpha $ entropy is defined by,
\begin{equation*}
\tilde{D}_{q }\left( \rho ||\sigma \right) =\frac{1}{q -1}\left( \mathrm{Tr}\rho ^{q }\sigma ^{1-q }-1\right),
\end{equation*}
for $q \in (0,2]$.
$\tilde{D}_{q }\left( \rho||\sigma \right)$ can also be reformulated as
\begin{equation*}
\tilde{D}_{q }\left( \rho ||\sigma \right) =\frac{1}{q -1}\left( \tilde{f}_{q}\left( \rho ,\sigma \right) -1\right),
\end{equation*}
where $\tilde{f}_{q }\left( \rho,\sigma \right) =\mathrm{Tr}\rho ^{q }\sigma ^{1-q }$.

Based on the Tsallis relative $\alpha $ entropy $\tilde{D}_{q }\left( \rho ||\sigma \right) $,
the coherence in the fixed reference basis $\left\{\left\vert j\right\rangle \right\} $ can be characterized by \cite{c26}
\begin{equation*}
\tilde{C}_{q }(\rho ) = \min_{\delta \in \mathcal{I}}\tilde{D}_{q }\left(\rho ||\delta \right).
\end{equation*}
Nevertheless, $\tilde{C}_{q}(\rho )$ violates the strong monotonicity condition of a coherence measure, even though it can
unambiguously distinguish the coherent states from the incoherent ones with the monotonicity.

%In Ref. \cite{c29}, a family of coherence quantifiers have been presented,
%\begin{equation*}
%C_{\alpha}(\rho)
%=\min_{\sigma\in \mathcal{I}}\frac{1}{\alpha-1}\left(f_{\alpha}^{\frac{1}{\alpha}}(\rho,\sigma)-1\right),
%\end{equation*}
%where $\alpha\in(0,2].$

In the following,  we define a generalized Tsallis relative operator entropy,
\begin{equation}\label{n7}
D_{q}(\rho||\sigma)
=\frac{1}{q-1}(f^{\frac{1}{q}}_q (\rho,\sigma)-1),
\end{equation}
where
\begin{equation}\label{n8}
f_{q}(\rho,\sigma)
=\mathrm{Tr}[\rho^{\frac{1}{2}}(\rho^{-\frac{1}{2}}\sigma \rho^{-\frac{1}{2}})^{1-q}\rho^{\frac{1}{2}}].
\end{equation}
$f_{q}(\rho,\sigma)$ has the following properties.

\begin{lemma}\label{lem2}
For any quantum states $\rho$ and $\sigma$ with $supp\,\rho\subseteq supp\,\sigma$, we have
$f_q(\Phi(\rho),\Phi(\sigma))\geq f_q(\rho,\sigma)$
for any CPTP map $\Phi$, where $q\in[0,1)$.
\end{lemma}

{\sf [Proof]}
Due to properties (VI), we get
\begin{equation}\label{qaz32}
\Phi[\rho^{\frac{1}{2}}(\rho^{-\frac{1}{2}}\sigma \rho^{-\frac{1}{2}})^{1-q}\rho^{\frac{1}{2}}-\rho] \leq \Phi(\rho^{\frac{1}{2}})(\Phi(\rho^{-\frac{1}{2}})\Phi(\sigma) \Phi( \rho^{-\frac{1}{2}}))^{1-q}\Phi(\rho^{\frac{1}{2}})-\Phi(\rho).
\end{equation}
For any CPTP map $\Phi$, we have
\begin{equation}\label{q32}
\mathrm{Tr}\left[\Phi[\rho^{\frac{1}{2}}(\rho^{-\frac{1}{2}}\sigma \rho^{-\frac{1}{2}})^{1-q}\rho^{\frac{1}{2}}-\rho] \right]
=\mathrm{Tr}[\rho^{\frac{1}{2}}(\rho^{-\frac{1}{2}}\sigma \rho^{-\frac{1}{2}})^{1-q}\rho^{\frac{1}{2}}]-\mathrm{Tr}\rho,
\end{equation}
and
\begin{equation}\label{q33}
\mathrm{Tr}\left[\Phi(\rho^{\frac{1}{2}})(\Phi(\rho^{-\frac{1}{2}})\Phi(\sigma) \Phi( \rho^{-\frac{1}{2}}))^{1-q}\Phi(\rho^{\frac{1}{2}})-\Phi(\rho)\right]
=f_q(\Phi(\rho),\Phi(\sigma))-\mathrm{Tr} \rho.
\end{equation}
According to (\ref {n8}), (\ref {qaz32}), (\ref {q32}) and (\ref {q33}), we get $f_{q}(\rho,\sigma) \leq f_{q}(\Phi(\rho),\Phi(\sigma))$.
$\Box$

Next, we give a lemma about the function $f_{q}(\rho,\sigma)$, which is important in deriving our main results.
Similar to the Lemma $1$ in Ref. \cite{sic}, we have

\begin{lemma}
Suppose both $\rho $ and $\sigma $ simultaneously undergo a TPCP map
$\Phi :=\left\{ K_{n}:\sum_{n}{K}_{n}^{\dagger }{K}_{n}= \mathcal{I}_{H}\right\} $
which transforms the states $\rho $ and $\sigma $ into the ensemble
$\left\{ p_{n},\rho_{n}\right\} $ and $\left\{q_{n},\sigma_{n}\right\} $,
respectively. We have
\begin{equation*}
f_{q}\left( \rho _{H},\delta _{H}\right) \leq \sum_{n} p_{n}^{q }q_{n}^{1-q}f_{q}\left( \rho _{n},\sigma _{n}\right).
\end{equation*}
\end{lemma}

{\sf [Proof]}
Any TPCP map can be achieved by unitary operations and local projection measurements on the composite system \cite{c2}.
Let $A$ be an auxiliary system. For a TPCP map $\Phi :=\left\{ K_{n}:\sum_{n}{K}_{n}^{\dagger }{K}_{n}= \mathcal{I}_{H}\right\}$,
we can always find a unitary operation $U_{H_A}$ and a set of projectors
$\left\{ \Pi _{n}^{A}=\left\vert n \right\rangle_{A}\left\langle n\right\vert \right\} $ such that
\begin{equation} \label{eq}
K_{n}\rho _{H}K_{n}^{\dagger }\otimes \Pi _{n}^{A}
=\left( \mathcal{I}_{H}\otimes \Pi _{n}^{A}\right) U_{H_A}\left( \rho_{H}\otimes \Pi _{0}^{A}\right) U_{H_A}^{\dagger }\left( \mathcal{I}_{H}\otimes \Pi _{n}^{A}\right).
\end{equation}
According to Lemma $1$ and the property (V), for any two states $\rho _{H}$ and $\sigma _{H}$ we have
\begin{equation*}
f_{q}\left( \rho _{H},\delta _{H}\right)
=f_{q}\left( U_{H_A}\left( \rho _{H}\otimes \Pi _{0}^{A}\right)
U_{H_A}^{\dagger },U_{H_A}\left( \sigma _{H}\otimes \Pi _{0}^{A}\right)
U_{H_A}^{\dagger }\right).
\end{equation*}

Denote $\rho _{H_f}=\Phi_{H_A}
\left[ U_{H_A}\left( \rho _{H}\otimes \Pi _{0}^{A}\right) U_{H_A}^{\dagger }
\right] $ and $\sigma _{H_f}=\Phi_{H_A}\left[ U_{H_A}\left( \sigma _{H}\otimes
\Pi _{0}^{A}\right) U_{H_A}^{\dagger }\right]$.
Due to Lemma $2$ we obtain
\begin{equation}\label{eq1r}
f_{q}\left( \rho _{H},\delta _{H}\right) \leq f_{q}\left( \rho _{H_f},\sigma _{Hf}\right) .
\end{equation}
Let the TPCP map be given by  $\Phi_{H_A}:=\left\{ \mathcal{I}_{H}\otimes \Pi_{n}^{A}\right\} $.
According to Eq. (\ref{eq}), $\rho _{H_f}$ and $\sigma _{H_f}$ can be replaced in Eq. (\ref{eq1r}),
respectively, by
\begin{equation*}
\rho _{H_f}\rightarrow \tilde{\rho}_{H_f}=\sum_{n}K_{n}\rho _{H}K_{n}^{\dagger }\otimes \Pi _{n}^{A}
\end{equation*}
and
\begin{equation*}
\sigma _{H_f}\rightarrow \tilde{\sigma}_{H_f}=\sum\limits_{n}K_{n}\sigma
_{H}K_{n}^{\dagger }\otimes \Pi _{n}^{A}.
\end{equation*}
Thus, we have
\begin{align*}
f_{q}\left( \rho _{H},\delta _{H}\right) \leq & f_{q}\left( \tilde{\rho}_{H_f},\tilde{\sigma}_{H_f}\right) \\\nonumber
= & \sum_{n}f_{q}\left( K_{n}\rho_{H}K_{n}^{\dagger }\otimes \Pi _{n}^{A}, K_{n}\sigma _{H}K_{n}^{\dagger}\otimes \Pi _{n}^{A}\right)  \\\nonumber
= & \sum_{n}f_{q}\left( K_{n}\rho_{H}K_{n}^{\dagger }, K_{n}\sigma _{H}K_{n}^{\dagger }\right)  \\\nonumber
= & \sum_{n}p_{n}^{q}q_{n}^{1-q}f_{q}\left( \rho _{n},\sigma _{n}\right),\\\nonumber
\end{align*}
which comletes the proof.
$\Box$

Based on the above results, we have the following main theorem.

\begin{theorem}
The coherence $C_{q}(\rho)$ of a quantum state $\rho$ given by
\begin{equation}\label{n9}
C_{q}(\rho)=\min_{\sigma\in \mathcal{I}}D_{q}(\rho||\sigma),
\end{equation}
defines a well-defined measure of coherence for $q\in(0,1)$.
\end{theorem}

{\sf Proof } From (\ref{n7}), (\ref{n8}) and (\ref{n9}), for $0< q<1$, we have
\begin{equation*}
C_{q}(\rho)=\min_{\sigma\in \mathcal{I}} \frac{1}{q-1}\left(f^{\frac{1}{q}}_q (\rho,\sigma)-1\right).
\end{equation*}
From Lemma 1, we have $C_{q}(\rho)\geq0$,
and $C_{q}(\rho)=0$ if and only if $\rho=\sigma.$

Next we prove that $C_{q}(\rho)$ satisfies $(C_3)$---strong monotonicity.
Let $\delta^{o}$ be the optimal incoherent state to the minimal value of $f_{q}(\rho,\delta)$,
i.e., $f_{q}(\rho,\delta^{o})=\max_{\delta \in \mathcal{I}} f_{q }(\rho ,\delta )$.
Let $\Phi= \{K_n\}$ be the incoherent selective quantum operations given by Kraus operators $\{K_n\}$,
with $\sum_{n}K_n^{\dagger}K_{n}={I}$, where ${I}$ is the identity operator on $H$.
Under the operation $\Phi$ on a state $\rho$, the post-measurement ensemble is given by $\{p_n, \rho_n\}$
with $p_n = \mathrm{Tr}K_n\rho K_n^{\dagger}$ and $\rho_n=K_n\rho K_n^{\dagger}/p_n$.
Hence the average coherence is
\begin{equation}\label{pfth11}
\sum_{n}p_n C_q(\rho_n)
=\min_{\delta _{n}\in \mathcal{I}}\frac{1}{q-1}\left(
\sum_{n}p_{n} f^\frac{1}{q}_q (\rho _{n},\delta _{n}) -1\right).
\end{equation}
Since the incoherent operation cannot generate coherence from an incoherent state,
for the optimal incoherent state $\delta^{o}$,
we have $\delta_n^o=K_n\delta^o K_n^{\dagger}/q_n\in \mathcal{I}$
with $q_n=\mathrm{Tr}K_n\delta^o K_n^{\dagger}$
for any incoherent operation $K_n $.
Due to $q\in(0,1)$ and $C_q(\rho)\geq 0$, $C_q(\rho)$ is the smallest when $f^{\frac{1}{q}}_{q }(\rho,\delta)$ is maximum.
Therefore, one immediately finds that
$\mathrm{max}_{\delta \in \mathrm{I}} f^{\frac{1}{q}}_{q }(\rho,\delta)\geq f^{\frac{1}{q}}_{q }(\rho_n,\delta_n^o)$. Therefore, Eq. (\ref{pfth11}) can be rewritten as
\begin{equation}\label{n11}
\sum_{n}p_n C_q(\rho_n)
\leq\frac{1}{q-1}\left(\sum_{n}p_{n}f^{\frac{1}{q}}_{q }\left( \rho _{n},\delta_n^o \right) -1\right).
\end{equation}

In addition, consider the H\"{o}lder inequality
\begin{equation*}
\sum_{k=0}^{d}a_k b_k\leq \left(\sum_{k=0}^{d} a^n_k\right)^\frac{1}{n}\left(\sum_{k=0}^{d} b^m_k \right)^\frac{1}{m},
\end{equation*}
for $ \frac{1}{n}+\frac{1}{m}=1 $ and $ n>1 $. The equality holds if and only if $\frac{a^n_k}{\sum_{k=0}^{d}a^n_k}=\frac{b^n_k}{\sum_{k=0}^{d}b^n_k}$,
and the inequality is reversed for $n\in (0,1)$. By using the H\"{o}lder inequality we obtain
\begin{equation}\label{n12}
\left[ \sum_{n}q_{n}\right] ^{1-q } \left[ \sum_{n}p_{n}f^{\frac{1}{q}}_{q } (\rho _{n},\delta _{n}^{o}) \right] ^{q }\geq
\sum_{n}p_{n}^{q }q_{n}^{1-q }f_{q} (\rho _{n},\delta_{n}^{o}) ,
\end{equation}
where $q \in (0,1)$.
Therefore, Eq. (\ref{n11}) becomes
\begin{align}\label{n13}
\sum_{n}p_n C_q(\rho_n) &\leq \frac{1}{q-1} \left(\sum_{n}p_{n}f^{\frac{1}{q}}_{q } ( \rho _{n},\delta_n^o) -1\right)\\\nonumber
&\leq\frac{1}{q-1}\left(\left[\sum_{n}p_{n}^{q }q_{n}^{1-q }f_{q }( \rho_{n},\delta_n^o)\right]^{\frac{1}{q}} -1\right)\\\nonumber
&\leq\frac{1}{q-1}\left(f^{\frac{1}{q}}_{q } (\rho,\delta^o ) -1\right)\\\nonumber
&= C_q(\rho),\nonumber
\end{align}
where the first inequality is due to Eq. (\ref{n11}), and
from Eq. (\ref{n12}) we get the second inequality.
The third inequality is due to Lemma $2$. Eq. (\ref{n13}) shows the strong monotonicity.
The monotonicity is directly given by the convexity of $C_q(\rho)$,
$C_q(\rho)\geq C_q(\sum_{n}p_n\rho_n)=C_q(\Phi(\rho))$.

Finally we prove that $C_{q}(\rho)$ satisfies condition $(C_5)$.
Suppose $\rho$ is block-diagonal in the reference basis $\{|j\rangle\}_{j=1}^{d}$,
$\rho=p_1\rho_1\oplus p_2\rho_2$
with $p_1\geq0,~ p_2\geq0$, $p_1+p_2=1,$ where $\rho_1$ and $\rho_2$ are density operators.
Let $\sigma=q_1\sigma_1\oplus q_2\sigma_2$
with $\sigma_1,~ \sigma_2$ the diagonal states having the same rows (columns) as $\rho_1,~ \rho_2$, respectively,
$q_1\geq0,~ q_2\geq0$, $q_1+q_2=1$.
It follows that
\begin{align}\label{mer}
&\max_{\sigma \in \mathcal{I}}\mathrm{Tr}\left(\rho^{\frac{1}{2}}(\rho^{-\frac{1}{2}}\sigma \rho^{-\frac{1}{2}})^{1-q}\rho^{\frac{1}{2}}\right)\\\nonumber
=&\max_{\sigma \in \mathcal{I}}\mathrm{Tr}\left\{(p_1\rho_1 + p_2\rho_2)^{\frac{1}{2}}\left [(p_1\rho_1 + p_2\rho_2)^{-\frac{1}{2}}(q_1\sigma_1 + q_2\sigma_2)(p_1\rho_1 + p_2\rho_2)^{-\frac{1}{2}}\right]^{1-q}(p_1\rho_1 + p_2\rho_2)^{\frac{1}{2}}\right\}\\\nonumber
=&\max_{\sigma \in \mathcal{I}}\mathrm{Tr}\left[(p^{\frac{1}{2}}_1\rho^{\frac{1}{2}}_1 + p^{\frac{1}{2}}_2\rho^{\frac{1}{2}}_2) (p^{-\frac{1}{2}}_1 q_1 p^{-\frac{1}{2}}_1 \rho^{-\frac{1}{2}}_1\sigma_1 \rho^{-\frac{1}{2}}_1 + p^{-\frac{1}{2}}_2 q_2 p^{-\frac{1}{2}}_2 \rho^{-\frac{1}{2}}_2 \sigma_2 \rho^{-\frac{1}{2}}_2)^{1-q} (p^{\frac{1}{2}}_1\rho^{\frac{1}{2}}_1 + p^{\frac{1}{2}}_2\rho^{\frac{1}{2}}_2)\right]\\\nonumber
=&\max_{\sigma \in \mathcal{I}}\mathrm{Tr}\left [p^{q}_1 q^{1-q}_1 \rho^{\frac{1}{2}}_1(\rho^{-\frac{1}{2}}_1\sigma_1 \rho^{-\frac{1}{2}}_1)^{1-q}\rho^{\frac{1}{2}}_1+
p^{q}_2 q^{1-q}_2 \rho^{\frac{1}{2}}_2 (\rho^{-\frac{1}{2}}_2\sigma_2 \rho^{-\frac{1}{2}}_2)^{1-q}\rho^{\frac{1}{2}}_2\right]\\\nonumber
=&\max_{q_{1},q_{2}}\{p^{q}_1 q^{1-q}_1 t_1 +p^{q}_2 q^{1-q}_2 t_2\},\nonumber
\end{align}
where we denoted
\begin{align*}
t_1 = \max_{\sigma_1} \rho^{\frac{1}{2}}_1\left(\rho^{-\frac{1}{2}}_1\sigma_1 \rho^{-\frac{1}{2}}_1\right)^{1-q}\rho^{\frac{1}{2}}_1,\\\nonumber
t_2 = \max_{\sigma_2} \rho^{\frac{1}{2}}_2\left(\rho^{-\frac{1}{2}}_2\sigma_2 \rho^{-\frac{1}{2}}_2\right)^{1-q}\rho^{\frac{1}{2}}_2.
\end{align*}
According to the H\"{o}lder inequality with $0<q<1$, we have
\begin{equation*}
p_1^{q} q_1^{1-q}t_1 + p_2^{q} q_2^{1-q}t_2
\leq \left(p_{1}t_{1}^{\frac{1}{q }}+p_{2}t_{2}^{\frac{1}{q }}\right)^{q },
\end{equation*}
where the equality holds if and only if $q_1=c p_1t^{\frac{1}{\alpha}}_1$
and $q_2=c p_2t^{\frac{1}{\alpha}}_2$ with $c=\left[p_1t^{\frac{1}{\alpha}}_1+ p_2t^{\frac{1}{\alpha}}_2\right]^{-1}$,
i.e, \begin{eqnarray}\label{min}
\max_ {q_1,q_2}\left(p_1^{q} q_1^{1-q}t_1 + p_2^{q} q_2^{1-q}t_2\right)
=\left(p_{1}t_{1}^{\frac{1}{q }}+p_{2}t_{2}^{\frac{1}{q }}\right)^{q }.
\end{eqnarray}
Combining (\ref{mer}) and (\ref{min}), we have
\begin{equation*}
\max_{\sigma\in \mathcal{I}}f^{\frac{1}{q}}_q (\rho,\sigma)=p_1 \max_{\sigma_1\in \mathcal{I}}f^{\frac{1}{q}}_q (\rho_1,\sigma_1)+p_2 \max_{\sigma_2\in \mathcal{I}}f^{\frac{1}{q}}_q (\rho_2,\sigma_2).
\end{equation*}
Thus, $C_{q}$ satisfies the additivity of coherence for block-diagonal states:
$C_{q}(p_1\rho_1\oplus p_1\rho_1)=p_1C_{q}(\rho_1)+p_2C_{q}(\rho_2).$
$\Box$

\section{Maximal coherence and several typical quantifiers}
We show that the maximal coherence of $C_{q}(\rho)$ ($q\in(0,1)$) can be attained by the maximally coherent states.
Based on the eigen-decomposition of a $d$-dimensional state $\rho=\sum_{j=1}^{d}\lambda_j|\varphi\rangle_j\langle\varphi|$,
where $\lambda_j$ and $|\varphi\rangle_j$ are the eigenvalues and eigenvectors of $\rho$, respectively.
We have
\begin{align*}
&f^{\frac{1}{q}}_{q}(\rho,\sigma)
=\left[ \mathrm{Tr}( \rho^{\frac{1}{2}}(\rho^{-\frac{1}{2}}\sigma \rho^{-\frac{1}{2}})^{1-q}\rho^{\frac{1}{2}})\right]^{\frac{1}{q}}\\\nonumber
=& \left\{ \mathrm{Tr}\left[\left(\sum_{k=1}^{d}\lambda_k|\varphi_k\rangle\langle\varphi_k|\right)^{\frac{1}{2}}
\left[\left(\sum_{j=1}^{d}\lambda_j|\varphi_j\rangle\langle\varphi_j|\right)^{-\frac{1}{2}}
\left(\sum_{i=1}^{d}\sigma_i|i\rangle\langle i|\right) \left(\sum_{j=1}^{d}\lambda_j|\varphi_j\rangle\langle\varphi_j|\right)^{-\frac{1}{2}}\right]^{1-q} \left(\sum_{k=1}^{d}\lambda_k|\varphi_k\rangle\langle\varphi_k|\right)^{\frac{1}{2}}\right]\right\}^{\frac{1}{q}}\\\nonumber
=& \left[\mathrm{Tr} \left(\sum_{k=1}^{d}\lambda_k|\varphi_k\rangle\langle\varphi_k|\left(\sum_{i,j=1}^{d} \lambda^{-1}_j \sigma_i |\langle\varphi_j|i\rangle|^{2} |\varphi_j\rangle\langle\varphi_j|\right)^{1-q}\right)\right]^{\frac{1}{q}}\\\nonumber
=& \left[ \mathrm{Tr} \left(\sum_{i,j=1}^{d} \lambda^{q}_j \left(\sigma_i |\langle\varphi_j|i\rangle|^{2}\right)^{1-q} |\varphi_j\rangle\langle\varphi_j|\right)\right]^{\frac{1}{q}}\\\nonumber
=& \left[\sum_{i,j=1}^{d}\lambda^{q}_j \left(\sigma_i |\langle\varphi_j|i\rangle|^{2}\right)^{1-q}\right]^{\frac{1}{q}}\\\nonumber
\geq & \sum_{i=1}^{d} \left[\sum_{j=1}^{d}\lambda^{q}_j \left(\sigma_i |\langle\varphi_j|i\rangle|^{2}\right)^{1-q}\right]^{\frac{1}{q}}\\\nonumber
\geq & d^{\frac{q-1}{q}} \left[\sum_{i,j=1}^{d}\lambda^{q}_j \left(\sigma_i |\langle\varphi_j|i\rangle|^{2}\right)^{1-q}\right]^{\frac{1}{q}}\\\nonumber
\geq & d^{\frac{q-1}{q}} \left(\sum_{i,j=1}^{d}\lambda_j \sigma_i |\langle\varphi_j|i\rangle|^{2}\right)^{\frac{1}{q}}\\\nonumber
\geq & d^{\frac{q-1}{q}},\\\nonumber
\end{align*}
where the first inequality is due to $(\sum_{i,j=1}^{d}a_i b_j)^{\frac{1}{q}}\geq \sum_{i=1}^{d} (\sum_{j=1}^{d} a_i b_j)^{\frac{1}{q}}$, with $a_i, b_i \geq 0 $.
The second inequality is due to that $\sum_{i=1}^{n}\lambda_i x_i^{p} \geq \left(\sum_{i=1}^{n}\lambda_i\right)^{1-p}
\left(\sum_{i=1}^{n}\lambda_ix_i\right)^{p}$, $p>1,$ with $x_i=\sum_{j=1}^{d}\lambda^{q}_j (\sigma_i |\langle\varphi_j|i\rangle|^{2})^{1-q} \geq0$, $\lambda_i=1$ $(i=1,2,...,n)$ and $p=\frac{1}{q}$.
The third inequality is due to $\sum_{k} a^{q}_k b^{1-q}_k \geq \sum_{k} a_k b_k,$ where $a_k, b_k \in(0,1)$.
Then one can easily find that the upper bound of the coherence can be attained by the maximally coherent states $\rho_d=|\varphi\rangle\langle\varphi|$
with $|\varphi\rangle=\frac{1}{\sqrt{d}}\sum_je^{i\phi_j}|j\rangle$. The corresponding coherence is given by
\begin{equation*}
C_{q }(\rho_d)=\frac{1}{q-1}(d^{\frac{q -1}{q }}-1).
\end{equation*}
$\Box$

$C_{q}(\rho)$ actually defines a family of coherence measures related to the Tsallis relative operator entropy.
Next, we introduce the special case of geometric coherence measures as the existing coherence measures,
and give detailed examples to illustrate the relationship between quantum coherence measures.
For $q=\frac{1}{2}$, one can also find that
\begin{equation}\label {12s}
C_{1/2}(\rho )=\min_{\sigma \in \mathcal{I}}2\left\{1-\left[ \mathrm{Tr}(\rho^{\frac{1}{2}}(\rho^{-\frac{1}{2}}\sigma \rho^{-\frac{1}{2}})^{\frac{1}{2}}\rho^{\frac{1}{2}})\right]^2 \right\},
\end{equation}
where
\begin{align} \label {qws}
f_{\frac{1}{2}}^{2}(\rho,\sigma) =& \left\{\mathrm{Tr}\left[\rho^{\frac{1}{2}}(\rho^{-\frac{1}{2}}\sigma \rho^{-\frac{1}{2}})^{\frac{1}{2}}\rho^{\frac{1}{2}}\right]\right\}^2 \\\nonumber
\leq & \left\{\mathrm{Tr}\left[\rho (\rho^{-\frac{1}{2}}\sigma \rho^{-\frac{1}{2}})\rho\right]^{\frac{1}{2}}\right\}^2 \\\nonumber
=& \left[\mathrm{Tr}(\rho^{\frac{1}{2}}\sigma \rho^{\frac{1}{2}})^{\frac{1}{2}}\right]^2,\nonumber
\end{align}
in which the inequality is due to the Araki-Lieb-Thirring inequality:
for matrixes $A, B\geq0$, $q\geq0$ and for $0\leq r\leq1$, the following inequality holds \cite{c32},
$tr(A^{r}B^{r}A^{r})^{q}\leq tr(ABA)^{rq}$.

We consider now the relationship between $C_{q}(\rho)$ and the geometric measure of quantum coherence.
The geometric measure of coherence $C_g (\rho)$ is defined by \cite{c31}
\begin{align}\label{z63}
C_g(\rho)= & 1-\mathrm{max}_{\sigma\in \mathcal{I}}F(\rho,\sigma)\\\nonumber
= & 1-\mathrm{max}_{\sigma\in \mathcal{I}}\left[\mathrm{Tr} (\rho^{1/2}\sigma\rho^{1/2})^{\frac{1}{2}}\right]^2,\\\nonumber
\end{align}
where $F(\rho,\sigma)$ is the Uhlmann fidelity of two density operators $\rho$ and $\sigma$.
From Eq. (\ref {qws}), we have that $F(\rho,\sigma)\geq f_{\frac{1}{2}}^{2}(\rho,\sigma)$.
Therefore, by the definition of geometric measure of coherence $C_g$, we get $C_{1/2}(\rho )\geq 2 C_g(\rho)$.

As an example, let us consider a single-qubit state,
$$
\rho=\frac{1}{2}(I_2+\sum_ic_i\sigma_i),
$$
where $I_2$ is the $2\times 2$ identity matrix and
$\sigma_i$ $(i=1,2,3)$ are Pauli matrices.
Suppose that $\sigma=\sum_ip_i|i\rangle\langle i|$ with $p_1+p_2=1$ and $0\leq p_1, p_2\leq1$.

For the single-qubit pure state $\rho$, with $\sum_ic_i^2=1$, one has
\begin{eqnarray}\label{rho}
\rho=\begin{pmatrix}
\frac{1+c_3}{2}& \frac{c_1-ic_2}{2} \\
\frac{c_1+ic_2}{2}&\frac{1-c_3}{2} \\
  \end{pmatrix}.\quad
\end{eqnarray}
The eigenvalues of $\rho$ are $0$ and $1$. Then, we obtain
\begin{align} \label{b63}
F(\rho,\sigma)= & \left[\mathrm{Tr} (\rho^{1/2}\sigma\rho^{1/2})^{\frac{1}{2}}\right]^2 \nonumber\\
= & \frac{1+c_3}{2}p_1 + \frac{1-c_3}{2}p_2.
\end{align}
Similar to the proof of (\ref{b63}), we obtain
\begin{align*}
f_{\frac{1}{2}}^{2}(\rho,\sigma) =& \left\{\mathrm{Tr}\left[\rho^{\frac{1}{2}}(\rho^{-\frac{1}{2}}\sigma \rho^{-\frac{1}{2}})^{\frac{1}{2}}\rho^{\frac{1}{2}}\right]\right\}^2 \\\nonumber
= & \frac{1+c_3}{2}p_1 + \frac{1-c_3}{2}p_2 . \\\nonumber
\end{align*}
Therefore, when $\rho$ is a single-qubit pure state, $F(\rho,\sigma)=f_{\frac{1}{2}}^{2}(\rho,\sigma)$, i.e., $C_{1/2}(\rho )= 2 C_g(\rho)$.

For the single-qubit state (\ref{rho}), with $\sum_ic_i^2<1$, the eigenvalues of $\rho$ are given by
\begin{align*}
\lambda_1= & \frac{1+\sqrt{c^2_1+c^2_2+c^2_3}}{2},\\\nonumber
\lambda_2= & \frac{1-\sqrt{c^2_1+c^2_2+c^2_3}}{2}.\\\nonumber
\end{align*}
Let $|\nu_1\rangle$ and $|\nu_2\rangle$ be the corresponding eigenvectors of $\rho$.
We have $\rho=\lambda_1 |\nu_1\rangle \langle \nu_1|+ \lambda_2 |\nu_2\rangle \langle \nu_2|$.
Combining (\ref{n8}) and (\ref{z63}), we obtain
\begin{align*}
F(\rho,\sigma)=& \lambda^{-\frac{1}{2}}_1 p^{\frac{1}{2}}_1 |\langle \nu_1|1\rangle|+ \lambda^{-\frac{1}{4}}_1 \lambda^ {-\frac{1}{4}}_2 p^{\frac{1}{2}}_1 |\langle \nu_2|1\rangle|+  \lambda^{-\frac{1}{4}}_1 \lambda^ {-\frac{1}{4}}_2 p^{\frac{1}{2}}_1 (\langle \nu_1|1\rangle \langle1|\nu_2\rangle)^{\frac{1}{2}} + \lambda^{-\frac{1}{2}}_1 p^{\frac{1}{2}}_2 |\langle \nu_1|2\rangle|+ \\\nonumber
& \lambda^{-\frac{1}{4}}_1 \lambda^ {-\frac{1}{4}}_2 p^{\frac{1}{2}}_2 |\langle \nu_2|2\rangle|+ \lambda^{-\frac{1}{2}}_2 p^{\frac{1}{2}}_1 |\langle \nu_2|1\rangle|+ \lambda^{-\frac{1}{4}}_1 \lambda^ {-\frac{1}{4}}_2 p^{\frac{1}{2}}_2 (\langle \nu_1|2\rangle \langle2|\nu_2\rangle)^{\frac{1}{2}}+ \lambda^{-\frac{1}{2}}_2 p^{\frac{1}{2}}_2 |\langle \nu_2|2\rangle|,\\\nonumber
f_{\frac{1}{2}}^{2}(\rho,\sigma)=& \lambda^{\frac{1}{2}}_1 p^{\frac{1}{2}}_1 |\langle \nu_1|1\rangle|+ \lambda^{-\frac{1}{4}}_1 \lambda^ {\frac{3}{4}}_2 p^{\frac{1}{2}}_1 |\langle \nu_2|1\rangle|+ \lambda^{\frac{3}{4}}_1 \lambda^ {-\frac{1}{4}}_2 p^{\frac{1}{2}}_1 (\langle \nu_1|1\rangle \langle1|\nu_2\rangle)^{\frac{1}{2}}+ \lambda^{\frac{1}{2}}_1 p^{\frac{1}{2}}_2 |\langle \nu_1|2\rangle|+
\\\nonumber
& \lambda^{-\frac{1}{4}}_1 \lambda^ {\frac{3}{4}}_2 p^{\frac{1}{2}}_2 |\langle \nu_2|2\rangle|+ \lambda^{\frac{1}{2}}_2 p^{\frac{1}{2}}_1 |\langle \nu_2|1\rangle|+ \lambda^{\frac{3}{4}}_1 \lambda^ {-\frac{1}{4}}_2 p^{\frac{1}{2}}_2 (\langle \nu_1|2\rangle \langle2|\nu_2\rangle)^{\frac{1}{2}}+ \lambda^{\frac{1}{2}}_2 p^{\frac{1}{2}}_2 |\langle \nu_2|2\rangle|.\\\nonumber
\end{align*}
Due to $0< \lambda_1, \lambda_2 <1$, we have $F(\rho,\sigma) > f_{\frac{1}{2}}^{2}(\rho,\sigma)$. Obviously $C_{1/2}(\rho )> 2 C_g(\rho)$.

\section{conclusion}
In summary, we have proposed four types of coherent measures $C_q (\rho)$ based on the Tsallis relative operator entropy.
It has been shown that these coherent measures meet all the necessary criteria for satisfactory coherence measures.
Moreover, the connections between $C_q (\rho)$ and the geometric measure of quantum coherence have been investigated.
Quantum coherence plays important roles in many quantum information processing. Our results may hight further researches on the characterization of quantum coherence.

\bigskip
\noindent{\bf Acknowledgments}\, \, This work is supported by NSFC under numbers 11765016, 11847209, 11675113, the GJJ170444, Key Project of Beijing Municipal Commission of Education (KZ201810028042), and Beijing Natural Science Foundation (Z190005), Academy for Multidisciplinary Studies, Capital Normal University, Shenzhen Institute for Quantum Science and Engineering, Southern University of Science and Technology, Shenzhen 518055, China (No. SIQSE202001).


\begin{thebibliography}{18}

\bibitem{c1} Scully, M.O., Zubairy, M.S.: Quantum Optics. (Canbrudge University Press, Cambridge, England, 1997)
\bibitem{c2} Nielsen, M.A., Chuang, I.L.: Quantum Computation and Quantum Information. (Canbrudge University Press, Cambridge, England, 2000)
\bibitem{c3} Rodriguez-Rosario, C.A., Frauenheim, T., Aspuru-Guzik. A.: Thermodynamics of quantum coherence. arXiv: 1308.1245
\bibitem{c4} Aberg, J.: Catalytic coherence. Phys. Rev. Lett. \textbf{113}, 150402 (2014)
\bibitem{c5} Narasimhachar, V.,  Gour, G.: Low-temperature thermodynamics with quantum coherence. arXiv: 1409.7740
\bibitem{c6} Cwiklinski, P., Studzinski, M., Horodecki, M., Oppenheim, J.: Towards fully quantum second laws of thermodynamics: limitations on the evolution of quantum coherences. Phys. Rev. Lett. \textbf{115}, 210403 (2015)
\bibitem{c7} Lostaglio, M., Jennings, D., Rudolph, T.: Description of quantum coherence in thermodynamic processes requires constraints beyond free energy. Nat. Commun. \textbf{6}, 6383 (2015)
\bibitem{c8} Lostaglio, M., Korzekwa, K., Jennings, D., Rudolph, T.: Quantum coherence, time-translation symmetry, and thermodynamics. Phys.
    Rev. X \textbf{5}, 021001 (2015)

\bibitem{nc8}Rastegin, A.E.: Coherence quantifiers from the viewpoint of their decreases in the measurement process.J. Phys. A: Math. Theor. \textbf{51}, 414011 (2018)

\bibitem{nc9}Bischof, F., Kampermann, H., Bruss, D.: Resource theory of coherence based on positive-operator-valued measures. Phys. Rev. Lett. \textbf{123}, 110402 (2019)

\bibitem{nc10}P.K. Dey, D. Chakraborty, P. Char, I. Chattopadhyay, D. Sarkar, Structure of POVM based resource theory of coherence. arXiv:1908.01882 [quant-ph]

\bibitem{c9} Baumgratz, T., Cramer, M., Plenio, M.B.: Quantifying coherence. Phys. Rev. Lett. \textbf{113}, 140401 (2014)
\bibitem{cq1} Li, C.M., Lambert, N., Chen, Y.N., Chen, G.Y., Nori, F.: Witnessing quantum coherence: from solid-state to biological systems. Sci. Rep. \textbf{2}, 885 (2012)
\bibitem{JX}   Xu, J.W.: Coherence measures based on sandwiched R$\acute{e}$nyi relative entropy.(arXiv:1808.04662v2)
\bibitem{c10} Yao, Y., Xiao, X., Ge, L., Sun, C.P.: Quantum coherence in multipartite systems. Phys. Rev. A \textbf{92}, 022112 (2015)
\bibitem{c11} Cheng. S., Hall, M.J.W.: Complementarity relations for quantum coherence. Phys. Rev. A \textbf{92}, 042101 (2015)
\bibitem{X2} Qi, X., Gao, T., Yan, F.: Measuring coherence with entanglement concurrence. J. Phys. A: Math. Theor. \textbf{50}, 285301 (2017)
\bibitem{c12} Winter, A., Yang, D.:  Operational resource theory of coherence. Phys. Rev. Lett. \textbf{116}, 120404 (2016)
\bibitem{PRA6} Luo, S.L.,  Zhang, Q.: Informational distance on quantum-state space. Phys. Rev. A \textbf{69}, 032106 (2004)
\bibitem{J1} Zhu, X.N., Jin, Z.X., Fei, S.M.: Quantifying quantum coherence based on the generalized $\alpha-z-$relative R$\acute{e}$nyi entropy. arXiv: 1905.01769
\bibitem{c13} Yuan, X., Zhou, H., Cao, Z., Ma, X.: Intrinsic randomness as a measure of quantum coherence. Phys. Rev. A \textbf{92}, 022124 (2015)
\bibitem{PR37} Yu, C.S.: Quantum coherence via skew information and its polygamy. Phys. Rev. A \textbf{95}, 042337 (2017)
\bibitem{Jin} Jin, Z.X.,  Fei, S. M.: Quantifying quantum coherence and non-classical correlation based on Hellinger distance. Phys. Rev. A \textbf{97},062342 (2018)
\bibitem{c14} Winter, A., Yang, D.: Operational resource theory of coherence. Phys. Rev. Lett. \textbf{116}, 120404 (2016)
\bibitem{c15} Bu, K.F., Singh, U., Fei, S.M., Pati, A.K.,  Wu, J.D.: Maximum relative entropy of coherence: an operational coherence measure. Phys. Rev. Lett. \textbf{119}, 150405 (2017)
\bibitem{c16} Xiong, C.H., Kumar, A.,  Wu, J.D.: Family of coherence measures and duality between quantum coherence
and path distinguishability. Phys. Rev. A \textbf{98}, 032324 (2018)
\bibitem{c17} Horodecki, R., Horodecki, P., Horodecki, M., Horodecki, K.: Quantum entanglement. Rev. Mod. Phys. \textbf{81}, 865 (2009)
\bibitem{c18} Vedral, V., Plenio, M.B., Rippin, M.A., Knight, P.L.: Quantifying entanglement. Phys. Rev. Lett. \textbf{78}, 2275 (1997)
\bibitem{c19} Vedral, V., Plenio, M.B.: Entanglement measures and purification procedures. Phys. Rev. A \textbf{57}, 1619 (1998)
\bibitem{c20} Plenio, M.B., Virmani, S., Papadopoulos, P.: Operator monotones, the reduction criterion and the relative entropy. J. Phys. A: Math. Gen. \textbf{33}, L193 (2000)
\bibitem{c94} Yu, X.D., Zhang, D.J., Xu, G.F., Tong, D.M.: Alternative framework for quantifying coherence. Phys.Rev. A \textbf{94}, 060302 (2016)
\bibitem{c21} Chitambar, E., Gour, G.: Critical examination of incoherent operations and a physically consistent resource theory of quantum coherence. Phys. Rev. Lett. \textbf{117}, 030401 (2016)
\bibitem{c22} Chitambar, E., Gour, G.: Comparison of incoherent operations and measures of coherence. Phys. Rev. A \textbf{95}, 019902 (2017)
\bibitem{c23} Liu, Z.W., Hu, X., Lloyd, S.: Resource destroying maps. Phys. Rev. Lett. \textbf{118}, 060502 (2017)
\bibitem{c24} Streltsov, A., Singh, U., Dhar, H.S., Bera, M.N., Adesso, G.: Measuring quantum coherence with entanglement. Phys. Rev. Lett. \textbf{115}, 020403 (2016)
\bibitem{c25} Rastegin, A.E.: Quantum-coherence quantifiers based on the Tsallis relative $\alpha$ entropies. Phys. Rev. A \textbf{93}, 032136 (2016)

\bibitem{nc25} Vershynina, A.: Quantum coherence, discord and correlation measures based on Tsallis relative entropy. Quantum Information and Computation \textbf{20}, No. 7, 553-569 (2020)

\bibitem{c26} Kollas, N.K.: Optimization-free measures of quantum resources. Phys. Rev. A 97, 062344 (2018)
\bibitem{c27} Yanagi, K., Kuriyama, K., Furuichi, S.: Generalized Shannon inequalities based on Tsallis relative operator entropy. Linear Algebra Appl. \textbf{394} (2005)
\bibitem{c29} Zhao, H.Q., Yu, C.S.: Remedying the strong monotonicity of the coherence measure in terms of the Tsallis relative $\alpha$ entropy. Sci. Rep. \textbf{8}, b299 (2018)
\bibitem{c31} Streltsov, A., Kampermann, H., Bru$\beta$, D.: Linking a distance measure of entanglement to its convex roof. New J. Phys. \textbf{12}, 123004 (2010)
\bibitem{clnx} Furuichi, S., Yanagi, K., Kuriyama, K.: A note on operator inequalities of Tsallis relative operator entropy. Linear Algebra Appl. \textbf{407} (2005)
\bibitem{c32} Audenaert, K.M.R.: On the Araki-Lieb-Thirring inequality. Int. J. Inf. Syst. Sci. \textbf{4}, 78 (2008)
\bibitem{c33} Chang, L.N., Luo, S.L.: Remedying the local ancilla problem with geometric discord. Phys. Rev. A \textbf{87}, 062303 (2013)
\bibitem{c34} Wigner, E.P., Yanase, M.M.: Information contents of distributions. Proc. Natl. Acad. Sci. \textbf{49}, 910 (1963)
\bibitem{c35} Lieb, E.H.: Convex trace functions and the Wigner-Yanase-Dyson conjecture. Adv. Math. \textbf{11}, 267 (1973)
\bibitem{aot} Fujii, J.I., Seo, Y.: Tsallis relative operator entropy with negative parameters. Adv. Oper. Theory \textbf{1}, (2016)
\bibitem{sic} Zhao, H.Q, Yu, C.S.: Coherence measure in terms of the Tsallis relative $\alpha$ entropy. Sci. Rep. \textbf{8}, 299 (2018)

\end{thebibliography}
\end{document}